\documentclass[pra,10pt,twocolumn,superscriptaddress,floatfix,showpacs]{revtex4-2}

\usepackage[utf8]{inputenc}  
\usepackage[T1]{fontenc}   
\usepackage[british]{babel} 
\usepackage[sc,osf]{mathpazo}
\usepackage[scaled=0.86]{berasans} 
\usepackage[colorlinks=true, allcolors=blue, urlcolor=blue]{hyperref} 
\usepackage{graphicx} 
\usepackage[babel]{microtype} 
\usepackage{amsmath,amssymb,amsthm,bm,amsfonts,mathrsfs,bbm} 

\usepackage{xspace} 
\usepackage{pgfplots}
\usepackage{xcolor,colortbl}
\usepackage{array}
\usepackage{bigstrut}

\newcommand{\tr}{\text{tr}}
\newcommand{\ket}[1]{| #1 \rangle}
\newcommand{\bra}[1]{\langle #1|}

\newcommand{\be}{\begin{equation}}
\newcommand{\ee}{\end{equation}}
\newcommand{\bea}{\begin{eqnarray}}
\newcommand{\eea}{\end{eqnarray}}
\newcommand{\bes}{\begin{equation*}}
\newcommand{\ees}{\end{equation*}}
\newcommand{\beas}{\begin{eqnarray*}}
	\newcommand{\eeas}{\end{eqnarray*}}

\newcommand{\x}{\mathrm{x}}
\newcommand{\ketbra}[1]{\ket{#1}\!\bra{#1}}

\def\x{\mathrm{x}}
\def\y{\mathrm{y}}

\def\I{\mathbbm{I}}

\def\tr{\mathrm{tr}}

\newtheorem*{thm*}{Theorem}

\newtheorem*{lem*}{Lemma}

\newtheorem*{lipschitzLem*}{Lemma \ref{lipschitz}}
\newtheorem*{lipschitzCubeLem*}{Lemma \ref{lipschitzCube}}
\newtheorem*{pgmNearlyOptimalThm*}{Theorem \ref{pgmNearlyOptimal}}

\begin{document}

\title{ Maximum confidence measurement for qubit states }

\author{Hanwool Lee }
\affiliation{School of Electrical Engineering, Korea Advanced Institute of Science and Technology (KAIST), 291 Daehak-ro, Yuseong-gu, Daejeon 34141, Republic of Korea }

\author{ Kieran Flatt}
\affiliation{School of Electrical Engineering, Korea Advanced Institute of Science and Technology (KAIST), 291 Daehak-ro, Yuseong-gu, Daejeon 34141, Republic of Korea }

\author{ Carles Roch i Carceller}
\affiliation{Department of Physics, Technical University of Denmark, 2800 Kongens Lyngby, Denmark}

\author{Jonatan Bohr Brask }
\affiliation{Department of Physics, Technical University of Denmark, 2800 Kongens Lyngby, Denmark}

\author{Joonwoo Bae}
\affiliation{School of Electrical Engineering, Korea Advanced Institute of Science and Technology (KAIST), 291 Daehak-ro, Yuseong-gu, Daejeon 34141, Republic of Korea }


\begin{abstract}
In quantum state discrimination, one aims to identify unknown states from a given ensemble by performing measurements. Different strategies such as minimum-error discrimination or unambiguous state identification find different optimal measurements. Maximum-confidence measurements (MCMs) maximise the confidence with which inputs can be identified given the measurement outcomes. This unifies a range of discrimination strategies including minimum-error and unambiguous state identification, which can be understood as limiting cases of MCM. In this work, we investigate MCMs for general ensembles of qubit states. We present a method for finding MCMs for qubit-state ensembles by exploiting their geometry, and apply it to several interesting cases, including ensembles two and four mixed states and ensembles of an arbitrary number of pure states. We also compare MCMs to minimum-error and unambiguous discrimination for qubits. Our results provide interpretations of various qubit measurements in terms of MCM and can be used to devise qubit protocols.

\end{abstract}


\maketitle

\section{Introduction} 

One fundamental difference between classical and quantum physics is that, while all information about the physical state of a quantum system is captured by its quantum state, such states are in general not perfectly distinguishable. Specifically, no measurement can perfectly discriminate non-orthogonal quantum states. This is closely related to other fundamental results in quantum mechanics such as the impossibility of perfectly copying quantum states \cite{wootters1982} and of faster-than-light signalling \cite{gisin1998}. The limits to discriminating between quantum states have numerous applications in quantum information science. Such limits are key to the security of quantum key distribution \cite{bennett1984,bennett1992}; near-optimal state discrimination enables approximate quantum error correction \cite{barnum2002}. They are also useful for operationally interpreting the differences between separable and entangled states \cite{PhysRevA.59.1070, Matthews:2009aa}, see also \cite{Spehner_2013, Spehner_2013q}. For further examples of the wide impact of quantum state discrimination, see the related reviews Refs.~\cite{chefles2000, bergou2004, bergou2007, barnett2009, bergou2010, bae2015, spehner2014}.

If it is impossible to perfectly discriminate quantum states, the natural thing to ask is precisely how well one can do. This, in turn, introduces the need for different figures of merit, corresponding to variations of the discrimination task. In general, the task consists in identifying states drawn from some ensemble, given a single copy of the state and prior knowledge of the possible states. Two well-studied cases are minimum-error and unambiguous state discrimination (MED and USD, respectively). In MED, one aims to minimise the probability that the state is misidentified while forbidding inconclusive outcomes  \cite{HELSTROM1967254, HELSTROM1968156, Helstrom:1969tb}. In USD, one instead enforces that the state is never misidentified, at the price of allowing for a non-zero inconclusive-outcome rate, which one then aims to minimise  \cite{ref:iva, ref:die, ref:per}. Both MED and USD are naturally formulated as statements about the conditional probabilities for observing certain outcomes, given that particular states were prepared.

Interestingly, distinct figures of merits in quantum state discrimination can be rephrased in terms of predictive and retrodictive formulations of quantum probabilities \cite{barnett2021}. Predictive probabilities are probabilities of future events conditioned on past events, which, in this context, are the probabilities of the outcomes conditioned on the input states. Retrodictive probabilities are probabilities of past events conditioned on future events occurring; here, this means the probabilities, conditioned on the observed outcomes, that particular input states were prepared. Predictive and retrodictive probabilities can be linked via Bayes’ theorem.

In this work, we focus on maximum-confidence discrimination, which is most naturally formulated in the retrodictive picture. The figure of merit here is the \textit{confidence}, defined as the conditional probability that an input was prepared given that the corresponding outcome was observed. A maximum confidence measurement (MCM) is a measurement strategy which achieves the best possible confidence. MCMs were introduced in Ref.~\cite{croke2006}. They unify the MED and USD settings of state discrimination. In particular, MCMs implement USD whenever USD is possible for the given ensemble and MED if a zero inconclusive rate is enforced and the maximum confidence considers an ensemble itself. In general, they make optimal use of detection events for guessing which states were prepared in the past \cite{PhysRevA.79.032323, PhysRevA.86.032314, PhysRevA.85.032312, PhysRevA.86.040303, PhysRevA.91.042338, PhysRevA.105.012609}.

We investigate MCMs for qubit states and determine general relations between a given ensemble and its MCM. We present a method for finding MCMs by exploiting the geometry of the Bloch sphere directly, without reference to the algebraic optimization problem, in a similar manner to geometric schemes for MED of $n$ qubit states \cite{PhysRevA.81.062304, PhysRevA.87.012334, PhysRevA.87.062302}. We then consider several particular ensembles of qubit states, derive their MCMs, and also compare to MED and USD.

The article is structured as follows. In Sec.~\ref{sec:medusd}, we start by briefly recalling the state discrimination problem in the simplest case of two pure states, and results for optimal MED and USD. In Sec.~\ref{sec:mcm}, we summarise MCMs.  In Sec.~\ref{sec:mcmq}, we formulate the problem of identifying an optimal MCM for qubits as a semidefinite program (SDP) and present optimality conditions. The relations between state ensembles and MCMs are found by exploiting the Bloch sphere geometry. In Sec.~\ref{sec:example}, various ensembles of qubit state ensembles are considered, and their MCMs are explicitly derived. We consider two mixed states, geometrically uniform states, tetrahedron states, and asymmetric states. In Sec.~\ref{sec:con} we conclude.

\section{MED and USD for two pure states}
\label{sec:medusd}

Let us consider the simplest non-trivial ensemble, consisting of two pure states, $|\psi_0\rangle$ and $| \psi_1 \rangle$, generated with {\it a priori} probabilities $q_0$ and $q_1$, respectively. A measurement device receives state $|\psi_{\x}\rangle$ with $\x\in\{0,1\}$, drawn from this ensemble, and provides an output $\y\in\{0,1,2\}$. The output can be understood as a guess for what input was prepared, i.e.~for the value of $\x$, with $\y=2$ denoting inconclusive outcomes. One can thus define an average error rate and an inconclusive rate, respectively, as
\begin{equation}
\label{eq:perr}
    \eta_{err} = q_0 \Pr(\y=1|\x=0) + q_1 \Pr(\y=0|\x=1) ,
\end{equation}
and
\begin{equation}
\label{eq:pinc}
    \eta_{inc} = q_0 \Pr(\y=2|\x=0) + q_1 \Pr(\y=2|\x=1) ,
\end{equation}
where $\Pr(\y|\x)$ denotes the conditional probability of observing outcome $\y$ given input $\x$. 

In MED, the goal is to minimise $\eta_{err}$ under the constraint that no inconclusive outcomes occur, i.e., $\Pr(\y=2|\x=0) = \Pr(\y=2|\x=1) = 0$. In this case, the minimal error rate is known as the Helstrom bound \cite{HELSTROM1967254, HELSTROM1968156, Helstrom:1969tb}
\begin{equation}
\label{eq:helstrom}
    \eta_{err} = \frac{1}{2} - \frac{1}{2} \| q_0 |\psi_0 \rangle \langle \psi_0 | - q_1 |\psi_1 \rangle \langle \psi_1 | \|_1 ,
\end{equation}
where $\| \cdot \|_1$ denotes the trace norm. 

This result applies to an arbitrary pair of quantum states and is found by a measurement with a construction as follows. As it is shown in Eq. (\ref{eq:helstrom}), the optimal measurement can be found in the support of given states $|\psi_0\rangle$ and $|\psi_1\rangle$. Then, two optimal positive-positive-operator-valued-measure (POVM) elements $M_0$ and $M_1$ are found as projectors with positive and negative eigenvalues of the operator $(q_0 |\psi_0 \rangle \langle \psi_0 | - q_1 |\psi_1 \rangle \langle \psi_1 |)$. 

One can also notice that, independently to a dimension of a Hilbert space where two states can be described, the two-state discrimination problem can be reduced to a two-dimensional space spanned by $|\psi_0\rangle$ and $|\psi_1\rangle$. In this sense, the two-state problem is equivalent to discrimination of two qubit states. Then, by referring to a Bloch sphere, an optimal measurement with POVM elements $M_0$ and $M_1$ can be found in a diameter of a half-plane due to the completeness, i.e., $M_0 +M_1= \mathbbm{I}$. The Helstrom bound in Eq. (\ref{eq:helstrom}) clarifies that the diameter should be parallel to the difference $(q_0 |\psi_0 \rangle \langle \psi_0 | - q_1 |\psi_1 \rangle \langle \psi_1 |)$.

In USD, on the other hand, the goal is to minimise $\eta_{inc}$ under the constraint that no errors occur, i.e., $\Pr(\y=1|\x=0) = \Pr(\y=0|\x=1) = 0$. In this case, the minimal inconclusive rate is
\begin{equation}
\label{eq:usdinc}    
    \eta_{inc} =2 \sqrt{q_0 q_1} |\langle \psi_0 | \psi_1\rangle | 
\end{equation}
If one hopes to be certain about which state was prepared, it suffices to rule out the other option. If one measurement outcome is $|\bar{\psi}_0\rangle$, such that $\langle \psi_0 |\bar{\psi}_0\rangle =0$, then that outcome can never occur when $|\psi_0\rangle$ is measured. This means that the prepared state must have been $|\psi_1\rangle$. The same holds for the other state, so that the POVM must include among its elements the two states orthogonal to those in the ensemble. A measurement consisting of just those outcomes, however, will not be complete, and so the POVM must be completed by a third element, which is the inconclusive one. Each of the elements must be weighted by constant factors and that associated with the third outcome determines the inconclusive rate. It is thus minimised. In this manner, the rate Eq. (\ref{eq:usdinc}) is attained. \cite{barnett2009}.

\section{ Maximum confidence measurement }
\label{sec:mcm}

We not turn to the more general case of discriminating between an arbitrary number of states. Let $S$ denote an ensemble of qubit states in which the states $\rho_{\x}$ are generated with {\it a priori} probabilities $q_{\x}$:
\bea
S = \{ q_{\x}, \rho_{\x}\}_{\x=0}^{n-1},~~\mathrm{and}~~\rho = \sum_{\x=0}^{n-1} q_{\x} \rho_{\x}. \label{eq:ens}
\eea
The most general measurement 
corresponds to an $n+1$-outcome positive-operator-valued-measure (POVM), denoted by $M = \{ M_{\y} \}_{\y=0}^n$, where outcome $n$ collects inconclusive events, while outcome $\y$ for $\y=0,\cdots,n-1$ denotes a guess that the input $\rho_{\y}$ was prepared.

Let $\rho_{\x}$ denote a state of particular interest in the ensemble. The probability that the correct state is identified is the confidence associated with the measurement \cite{croke2006}, 
\bea
C(\x) &:=& \Pr (\rho_x | M_x ) \nonumber \\
&=&\frac{\Pr(\rho_x)\Pr(M_x|\rho_x)}{\Pr(M_x)} = \frac{q_{\x}\tr[\rho_{\x} M_{\x}]}{\tr[\rho M_{\x}]} , \label{eq:conf}
\eea
where Bayes' rule is applied and $\Pr (M_{\x} | \rho_{\x})$ is the probability that the outcome associated with $M_{\x}$ is triggered by the state $\rho_{\x}$. For example, $C(\x)=1$ for some $\x$ signifies unambiguous identification of the state $\rho_{\x}$ by a detection event on $M_{\x}$. Given a detection event, a state $\rho_{\x}$ is verified with certainty. Unambiguous discrimination of quantum states is achieved when $C(\x)=1$ for all $\x=0,\cdots,n-1$. 

The confidence in Eq. (\ref{eq:conf}) can be maximized by optimizing over each POVM element according to
\bea
\max C(\x) = \max_{M_{\x}} \frac{q_{\x}\tr[\rho_{\x} M_{\x}]}{\tr[\rho M_{\x}]} , \label{eq:maxconf}
\eea
where $0\leq M_{\x} \leq 1$. A valid POVM, which attains the optimum for all $\x$, can always be obtained by rescaling the $M_{\x}$ and including one additional element $M_{\phi}$ which collects inconclusive outcomes. Such a measurement is called an MCM. In general, we have $M_{\phi}\neq 0$.

As mentioned, when unambiguous discrimination is possible for an ensemble, the MCM is identical with the measurement giving unambiguous discrimination. An MCM for an ensemble of two pure states, for instance, will identify each state with perfect confidence. Note, however, that an MCM can be introduced for ensembles for which unambiguous discrimination is impossible, such as three qubit states. 

One may consider the maximum confidence for an ensemble itself: writing by $p_{\x}$ as the probability that a detector $M_{\x}$ shows a detection event, i.e., $p_{\x} = \tr[\rho M_{\x}]$, the maximization 
\bea
\max \sum_{\x=0}^{n-1} p_{\x} C(\x) 
\eea
over a complete measurement equals to the highest success probability in minimum-error state discrimination \cite{barnett2009}. We remark that an MCM provides a unifying picture of different figures of merits in quantum state discrimination. 

Note that the maximization in Eq. (\ref{eq:maxconf}) is computationally feasible \cite{croke2006}. One can apply the transformation  
\bea
\widetilde{\rho}_{\x}=\sqrt{\rho}^{-1} q_{\x} \rho_{\x} \sqrt{\rho}^{-1} ~~\mathrm{where}~~\rho = \sum_{j} q_j \rho_j \label{eq:trans}
\eea  
in order to rewrite the optimization problem in Eq. (\ref{eq:maxconf}) as
\bea
 \max C(\x) = \max_{Q_{\x}\geq 0,~ \tr[Q_{\x}]=1} \tr[\widetilde{\rho}_{\x} Q_{\x} ].  \label{eq:max1}
\eea
It is not difficult to see that the maximum confidence above corresponds to the operator norm 
\bea
\max C(\x) = \|\sqrt{\rho}^{-1} q_{\x} \rho_{\x} \sqrt{\rho}^{-1} \|_{op} \label{eq:max11} 
\eea
where $\|\cdot \|_{op}$ denotes the operator norm $\| A \|_{op} = \sup_{\| v \|=1} \| Av \|$. Once an optimal operator in Eq. (\ref{eq:max1}), denoted by $Q_{\x}^{*}$, is obtained, an optimal POVM element $M_{\x}^{*}$ is found as
\bea
M_{\x}^{*} =c_{\x} \sqrt{\rho}^{-1} Q_{\x}^{*} \sqrt{\rho}^{-1}. 
\eea
for some constant $c_{\x}>0$. Note that $\{c_{\x}\}$ may be chosen such that $\sum_{\x} M_{\x}^{*} \leq \mathbbm{I}$.

\section{ MCM for Qubit States}
\label{sec:mcmq}

In this section, we approach the maximum confidence in Eq. (\ref{eq:maxconf}) from the point of view of convex optimization. We first show a semidefinite program (SDP) for the optimization problem and then analyze the optimality conditions in order to show that a general structure relates the states to their MCM. 

\subsection{Convex optimization}

We begin with the maximization problem in Eq. (\ref{eq:max1}), which is linear with respect to a state of interest. The optimization problem can be written as an SDP as follows,
\bea
p^* = \max &~~& \tr[\widetilde{\rho}_{\x} Q_{\x}] \label{eq:primal} \\
\mathrm{subject~to} &~~& Q_{\x}\geq0 ~\mathrm{and} ~\tr[Q_{\x}]=1. \nonumber
\eea 
Its dual problem is found by constructing the Lagrangian 
\bea
\mathcal{L}(Q_{\x},\lambda_{\x}, Z_{\x}) & =& \tr[\widetilde{\rho}_{\x} Q_{\x}] + \lambda_{\x}(1-\tr[Q_{\x}])+\tr[Q_{\x}Z_{\x}]~~~~~~
\eea
where $Z_{\x}\geq 0$ and $\lambda_{\x}$ are dual variables. Maximizing this Lagrangian gives the dual function 
\bea
g(\lambda_{\x}, Z_{\x})& = &\max_{Q_{\x}} \mathcal{L} (Q_{\x},\lambda_{\x}, Z_{\x}) \nonumber \\
& = & \lambda_{\x} + \max_{Q_{\x}} \tr[(\widetilde{\rho}_{\x} - \lambda_{\x} I + Z_{\x} ) Q_{\x}]. 
\eea
It can be seen that the function $g(\lambda_{\x}, Z_{\x})$ does not converge if $\tilde{\rho}_{\x}- \lambda_{\x} I + Z_{\x} \neq 0$. Therefore, the optimal dual parameters $Z_{\x}^*$ and $\lambda_{\x}^*$ satisfy the condition
\bea
\widetilde{\rho}_{\x} - \lambda_{\x}^* \I+ Z_{\x}^*=0, \label{eq:ls1}
\eea
called the Lagrangian stability. We note that the primal problem is feasible. The next condition that optimal parameters satisfy is called the complementary slackness, given as 
\bea
\tr[Z_{\x}^* Q_{\x}^*] =0. \label{eq:cs1}
\eea
The dual problem is then obtained as 
\bea
d^* = \min &~~& \lambda_{\x} \label{eq:dual} \\
\mathrm{subject~to} &~~& \lambda_{\x} \rho - q_{\x}\rho_{\x} \geq 0. \nonumber
\eea
Since both problems are feasible, one can find the maximum confidence from both primal and dual problems above, $p^* = d^* = \max C(\x)$. 

The linear complementarity problem (LCP) approach may be used to understand the convex optimization problem's structure \cite{Cottle2009}. Technically speaking, while an SDP, either primal or dual problem, is formed with inequalities, an LCP directly analyzes the optimality conditions, which are given in terms of equalities. Those primal and dual parameters satisfying the equalities automatically find an optimal solution.

We are now in a position to derive the optimality conditions in terms of an ensemble $\rho$ and state of interest $\rho_{\x}$. From the transformation in Eq. (\ref{eq:trans}), we introduce new parameters $r_{\x} > 0$ and a state $\sigma_{\x}$ such that $r_{\x} \sigma_{\x} = \sqrt{\rho} Z_{\x}^* \sqrt{\rho}$, so that the optimality conditions can be rewritten as
\bea
\mathrm{Lagrangian ~stability~}&:& \lambda_{\x} \rho = q_{\x} \rho_{\x} + r_{\x} \sigma_{\x} \label{eq:ls} \\
\mathrm{Complementary ~slackness~} &:& r_{\x} \tr [\sigma_{\x} M_{\x} ] = 0 . \label{eq:cs}
\eea
Since both primal and dual problems are feasible, those primal and dual parameters satisfying Eqs. (\ref{eq:ls}) and (\ref{eq:cs}) automatically pinpoint the optimization problem's solution. Once dual parameters are found from Eq. (\ref{eq:ls}), the optimal POVM element is characterized by Eq. (\ref{eq:cs}). Note that an optimal POVM element is found by the equalities given in the optimality conditions.

\subsection{MCM for qubit states}

We now investigate the optimality conditions for qubit states and show how one can solve the optimization problem directly. Both the maximum confidence and optimal POVM elements can be found. Let us begin with the condition in Eq. (\ref{eq:cs}). The product of a complementary state $\sigma_{\x}$ and an optimal POVM must be zero. Since the optimal measurement satisfies $M_{\x}^{*}\neq 0$ for $\forall \x=0,\cdots, n-1$, it holds that both $\sigma_{\x}$ and $M_{\x}^*$ must be rank-one and orthogonal with each other. 

Let us consider the Lagrangian stability in Eq. (\ref{eq:ls}), which can be rewritten for all $\x=0,\ldots,n-1$ as 
\bea
\rho = \mu_{\x} \rho_{\x} + (1-\mu_{\x}) \sigma_{\x},~\mathrm{where}~\mu_{\x} =\frac{q_{\x}}{\lambda_{\x}}. \label{eq:1}
 \eea
Note that decompositions above for qubit states have been also obtained in Ref. \cite{PhysRevA.105.012609}. MCMs can be computed analytically from this relation, which implies
\bea
\tr[ \sigma_{\x}^2] = \frac{1}{(1-\mu_{\x})^2} \tr [(\rho-\mu_{\x}\rho_{\x})^2]. ~\label{eq:relation}
\eea
For qubit cases, the left-hand side is given by $1$ since the complementary state is rank-one. It is straightforward to find the value $\mu_{\x}$ as follows. Suppose that the state of interest $\rho_{\x}$ is pure, i.e., $\tr[\rho_{\x}^2]=1$. We then have 
\bea
\mu_{\x} = \frac{1-\tr[\rho^2]}{2(1 - \tr [\rho\rho_{\x}] ) }, \label{eq:f1}
\eea
which can be computed from an ensemble $\rho$ and a state of interest $\rho_{\x}$. 

When a state $\rho_{\x}$ is not pure, we have
\bea
\mu_{\x} & = & \frac{(1- \tr [ \rho \rho_{\x}] ) - \mathrm{Det}(\rho,\rho_{\x})}{1-\tr [ \rho_{\x}^2]} ~~\mathrm{where}~ \label{eq:f2}\\
\mathrm{Det}(\rho,\rho_{\x}) & = & [(1-\tr [ \rho \rho_{\x}] )^2 -(1-\tr [ \rho^2 ])(1-\tr [ \rho_{\x}^2] )]^{1/2}. \nonumber
\eea
The maximum confidence is obtained as
\bea
\max C(\x) = \lambda_{\x}^* = \frac{q_{\x}}{\mu_{\x}}, \label{eq:Cmu}
\eea
when the state is prepared with {\it a priori} probability $q_{\x}$. We have therefore shown how to compute the maximum confidence for a state of interest. Once $\mu_{\x}$ is found as above, one can find the complementary state $\sigma_{\x}$ in Eq. (\ref{eq:1}), from which the optimal POVM element is also found.

\subsubsection*{Example: $N$ qubit pure states}

To illustrate our approach, let us consider an ensemble of $N$ arbitrary pure states:
\bea
&& \left\{\ket{\psi_j}\right\}_{j=0}^{N-1}, ~\mathrm{where } ~\ket{\psi_0} = \ket{0}~\mathrm{and} \\
&&  \ket{\psi_j} = \cos\frac{\theta_j}{2}\ket{0} +   e^{i \phi_j} \sin\frac{\theta_j}{2}|1\rangle ~\mathrm{for}~j=1,\cdots,N-1. \nonumber
\eea
Note that the angles ($\theta_j$, $\phi_j$) are arbitrary and the state of interest is denoted by $\ket{\psi_0}$. One can compute the maximum confidence as
\bea
    \max C(0)=\frac{2\left(1-\tr\left[\rho_0\rho_{M}\right]\right)}{N+1-(N-1)\tr\left[\rho_{M}^{2}\right]-2\tr\left[\rho_0\rho_{M}\right]}.~~~ \label{eq:maxconN}
\eea
where $\rho_M$ is an equally weighted mixture of $N- 1$ states $|\psi_j\rangle$ for $j=1,\cdots, N-1$. It is seen that the maximum confidence depends on two parameters: the purity of an ensemble $\rho_M$ and the fidelity between $\rho_M$ and $\rho_0$. 

In addition, as shown in Refs. \cite{flatt2021contextual, carceller2021quantum}, the maximum confidence is closely related to the outcome rate, the probability that a detection event occurs, denoted by $\eta_0 = \tr[\rho M_0]$. Here, the outcome rate is upper-bounded by 
\bea
\eta_+ 
&=&1+\frac{\mu_0 \tr[\rho \rho_0]-\tr[\rho^2]}{1-\mu_0} 
\eea
where $\mu_0= ({1-\tr[\rho^2]}) / ({2(1-\tr[\rho \rho_0])}) $, see Eq. (\ref{eq:f1}).

It's worth emphasizing that any $N$ state discrimination problem within the MCM framework can be turned into a $2$ state discrimination problem. Since an MCM only focuses on one state of interest ($\rho_0$), the rest can be collected in a mixture $\rho_ {M}$. Maximum confidence can be straight computed with \eqref{eq:maxconN}, which is equivalent to \eqref{eq:Cmu} for equiprobable preparations. \\

\begin{figure}[]
  \centering
  \includegraphics[scale=0.26]{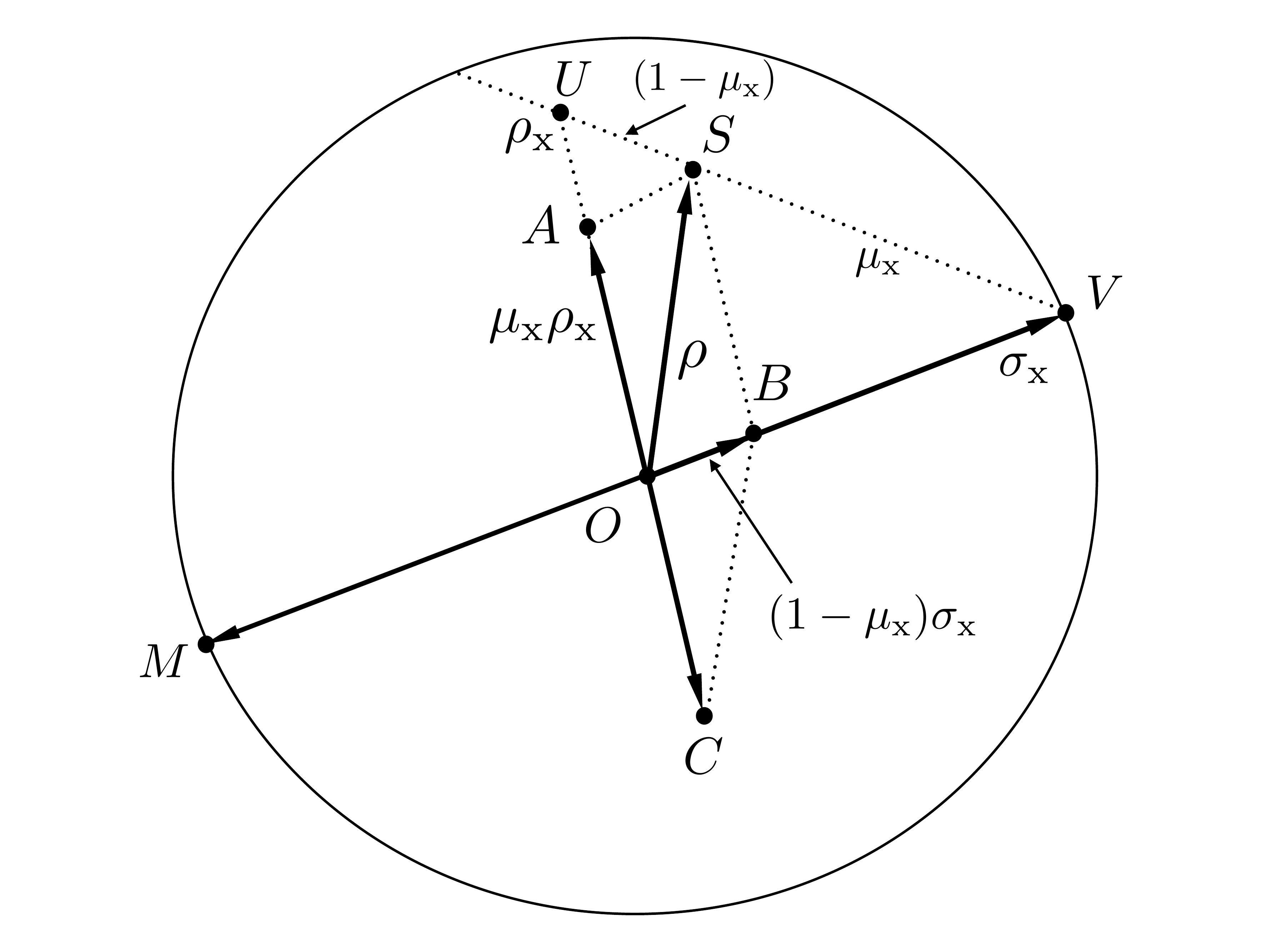}
  \caption{ The geometry of MCM for qubit states is shown on the Bloch sphere. The arrows represent Bloch vectors. For instance, Bloch vectors $OM$ and $OV$ on the sphere, i.e., pure states, denote orthogonal qubit states. An ensemble $\rho$ and a state of interest $\rho_{\x}$ correspond to $OS$ and $OU$ respectively. A complementary state $\sigma_{\x}$ is pure, i.e., lies on the sphere. Since $\rho$ is a convex combination of $\rho_{\x}$ and a complementary state $\sigma_{\x}$, see Eq. (\ref{eq:1}), the state $\sigma_{\x}$ is immediately obtained as $OV$ by extending $US$. An optimal POVM element corresponds to $OM$. It holds that $OA + OB = OS$ and $OS + OC = OB \propto OV$. }
  \label{fig:scheme}
\end{figure}

\subsection{Geometry of qubit states and an MCM}

The general structure of qubit states and MCMs can be depicted on the Bloch sphere. We here analyze the optimality condition geometrically and present the structure. We also show forms of the maximum confidence different to Eq. (\ref{eq:Cmu}). 

Let us refer to Fig. \ref{fig:scheme}. Note that the natural distance measure in the Bloch sphere is given by the Hilbert-Schmidt norm, which turns out to be proportional to the trace norm for qubit cases \cite{bae2013}, i.e.,
\bea
\sqrt{2} d_{HS} (\rho,\sigma) =   \|\rho-\sigma \|_1 
\eea
where $d_{HS}(\rho,\sigma) = \sqrt{ \tr[ (\rho-\sigma)^2]}$. For instance, the trace norm between two orthogonal qubit states equals to $2$ and the Hilbert-Schmidt distance by $\sqrt{2}$. Thus, one can consider two measures interchangeably in the Bloch sphere and relate them by a factor of $\sqrt{2}$. 

We begin by interpreting Eq. (\ref{eq:1}): an ensemble $\rho$ is given as a convex mixture of a state of interest $\rho_{\x}$ and its complementary one $\sigma_{\x}$. This means that the Bloch vector of a state $\rho$ lies on a line connecting two Bloch vectors of two states $\rho_{\x}$ and $\sigma_{\x}$. It also implies that the Bloch vector of a state $\sigma_{\x}$ can be found on a line connecting those of two states $\rho$ and $\rho_{\x}$. Let us recall from the optimality condition in Eq. (\ref{eq:cs}) that a complementary state $\sigma_{\x}$ must be rank-one on the Bloch sphere. Therefore, one can find a complementary state $\sigma_{\x}$ on the surface at which the line connecting two known states $\rho$ and $\rho_{\x}$ meets, see Fig. \ref{fig:scheme}. Once a complementary state is found, an optimal POVM element is obtained as the orthogonal complement, 
\bea
M_{\x}^* \propto \sigma_{\x}^{\perp}. 
\eea
Both operators $M_{\x}^*$ and $\sigma_{\x}$ are rank-one. 

Let us also explain the relations between the states and MCM, as shown in Fig. \ref{fig:scheme}. Given states $\rho$ and $\rho_{\x}$, displayed as $OS$ and $OU$ respectively, an optimal measurement is found as $OM$ that is orthogonal to $OV$ obtained on the sphere by extending $US$. The Bloch vector of a complementary state that corresponds to $OV$ can be found as follows.

Throughout, let $\vec{n}(\tau)$ denote the Bloch vector of a qubit state $\tau$. Then, a vector $\vec{r}$ lying on a line defined by $US$ is given by
\bea
\vec{r}=(\vec{n}(\rho)-\vec{n}(\rho_{\x}))t+\vec{n}(\rho) \label{eq:lineq}
\eea
for some $t \geq 0$. The complementary state's Bloch vector $\vec{r}_{\x}$ is found with $t_{\x}$, by which $\| \vec{r} \| =1$.

From the convex combination in Eq. (\ref{eq:1}), it holds that 
\bea
\frac{\|\rho-\rho_{x} \|_1 }{ \| \rho -\sigma_{\x}\|_1} = \frac{1-\mu_{\x} }{ \mu_{\x}}. 
\eea
From the relation above, it is straightforward to find $\mu_{\x} = \| \rho-\sigma_{\x} \|_1 / \| \rho_{\x} - \sigma_{\x} \|_1$, so that
\bea
\max C(\x) &=& \frac{q_{\x}\| \rho_{\x} -\sigma_{\x}\|_1 }{\| \rho - \sigma_{\x} \|_1}.~~~~~~ \label{eq:r}
\eea
From Eq. (\ref{eq:lineq}), one can compute the maximum confidence above in terms of the Bloch vector. It follows that
\bea
 \max C(\x) = q_{\x} \left(1+ \frac{1}{t_{\x}} \right) \label{eq:ct}
\eea
where $t_{\x}$ is found from the constraint $\| \vec{r}\|=1$. 

We have therefore shown that a complementary state can be directly found by exploiting the qubit state geometry, as well as an MCM. In summary, the maximum confidence for qubit states can be written in the various forms in Eqs. (\ref{eq:Cmu}), (\ref{eq:r}) and (\ref{eq:ct}).

\subsection{ Minimizing the probability of inconclusive outcomes }\label{subsec:inc}

Having found POVM elements for an MCM, let us consider the probability inconclusive outcomes. As it is mentioned, it is clear that a POVM element in an MCM is rank-one: for an ensemble in Eq. (\ref{eq:ens}), let 
\bea
M_{\x} = a_{\x} \Pi_{\x} \label{eq:prod}
\eea
denote a POVM element for each state where $a_{\x}$ is a non-negative constant and $\Pi_{\x}$ a rank-one projector. The projectors $\{\Pi_{\x}\}_{\x=0}^{N-1}$ are immediately obtained such that they perform an MCM. Then, a set of constants $\{ a_{\x}\}_{\x=0}^{N-1}$ is chosen to find the probability of inconclusive outcomes, for which the POVM element is denoted by
\bea
M_{\phi} = \mathbbm{I} - \sum_{\x} a_{\x} \Pi_{\x} \label{eq:mphi}
\eea
so that its probability is given by $p_{inc} =  \tr[\rho M_{\phi}]$. 

Remarks are in order. Firstly, an MCM for an ensemble (see Eq. (\ref{eq:ens})) varies by choosing different values $\{ a_{\x}  \}_{\x=0}^{N-1}$, for all of which an MCM holds true. This immediately concludes that an MCM for an ensemble is not unique. Secondly, if the convex hull of POVM elements $\{ \Pi_{\x}\}$ performing an MCM contains the identity, i.e., $\{ a_{\x} \}_{\x=0}^{N-1}$ can be chosen such that $\sum_{\x=0}^{N-1} a_{\x} \Pi_{\x} = \mathbbm{I}$, one can find an MCM that is also complete. Consequently, an inconclusive outcome does not occur, since $M_{\phi}=0$ and $p_{inc} =0$.  \\

 {\bf Remark.} Let $\{\Pi_{\x} \}$ denote a set of rank-one projectors and suppose that their convex hull contains the identity. Then, an MCM with the projectors, $\{a_{\x} \Pi_{\x} \}$, can be constructed such that an inconclusive outcome does not occur.   \\

Thirdly, if the convex hull of POVM elements $\{ \Pi_{\x}\}$ does not contain the identity, an optimization problem is introduced to minimize the probability of inconclusive outcomes. From a POVM in Eq. (\ref{eq:mphi}), the problem is defined as follows. 
\bea
Q &=& \min ~~\tr[\rho M_{\phi}] \label{eq:mphi1} \\
\mathrm{subject~to~} && a_{\x}\geq0,~\forall \x,~~\mathrm{and}~M_{\phi} \geq 0. \nonumber
\eea
The optimization problem may be approached by a Lagrangian,
\bea
{\mathcal{L} =  \tr[\rho M_{\phi}]  -\sum_{\x}v_{\x}a_{\x} - \tr[K M_{\phi} ]} \nonumber 
\eea
where $K\geq 0$ and $v_{\x}\geq0$ are dual parameters. The optimality conditions contain the Lagrangian stability, i.e, $\partial \mathcal{L} / \partial a_{\x}=0$, for all $\x$,
\bea
\tr[\rho \Pi_{\x}] + v_{\x} -\tr [\rho \Pi_{\x}] = 0, \label{eq:lsmphi}
\eea 
and the complementary slackness,
\bea
v_{\x} a_{\x} =0,~~\mathrm{and}~~{\tr}KM_{\phi}=0. \label{eq:mphics}
\eea
The optimization problem works for an arbitrary ensemble of quantum states. In what follows, let us rewrite the problem specifically for qubit states.  

It is straightforward to find that Eq. (\ref{eq:mphics}) implies $M_{\phi}$ is rank-one for qubit states. Hence, it holds that
\bea
&&\big( \frac{M_{\phi} }{\tr[M_{\phi} ]} \big)^2= \frac{M_{\phi} }{\tr[M_{\phi} ]} \nonumber
\eea
which is equivalent to, from Eq. (\ref{eq:mphi}),
\bea
 1-\sum_{\x} a_{\x} + \frac{1}{2} \sum_{\x,\y} (1- \tr[\Pi_{\x} \Pi_{\y} ]) a_{\x}a_{\y} = 0.  \label{eq:mphir1}
\eea
In addition, also from Eq. (\ref{eq:mphi}), we have $\tr [M_{\phi} ] \geq 0$, meaning that $2-\sum_{\x}a_{\x} \geq 0$. With the constraints, the optimization problem in Eq. (\ref{eq:mphi1}) for qubit states can be written as follows,
\bea
Q &=& \min ~~ \tr[\rho (1-\sum_{\x}a_{\x} \Pi_{\x} )] \label{eq:mphi2} \\
\mathrm{subject~to~} &&\forall \x,~ a_{\x}\geq0,~~2-\sum_{\x} a_{\x} \geq0,~~\mathrm{and}  \nonumber\\
&&  1-\sum_{\x} a_{\x} + \frac{1}{2} \sum_{\x,\y} (1- \tr[\Pi_{\x} \Pi_{\y} ]) a_{\x}a_{\y} = 0. \nonumber
\eea
The probability of inconclusive outcomes for an MCM of qubit states can be generally obtained by solving the optimization problem. We reiterate that, once a set of projectors for an MCM $\{\Pi_{\x} \}_{\x=1}^{N-1}$ is obtained, the optimization problem above finds a set of optimal coefficients $\{a_{\x} \}_{\x=1}^{N-1}$ to minimize the probability of inconclusive outcomes.

\section{ Various Qubit States}
\label{sec:example}

Let us apply the geometric structure of MCM to various ensembles of qubit states. We show how states and their MCM are related to each other. We also compare MCMs for qubit states to measurements for unambiguous and minimum-error discrimination.

\begin{figure}[]
  \centering
  \includegraphics[scale=0.26]{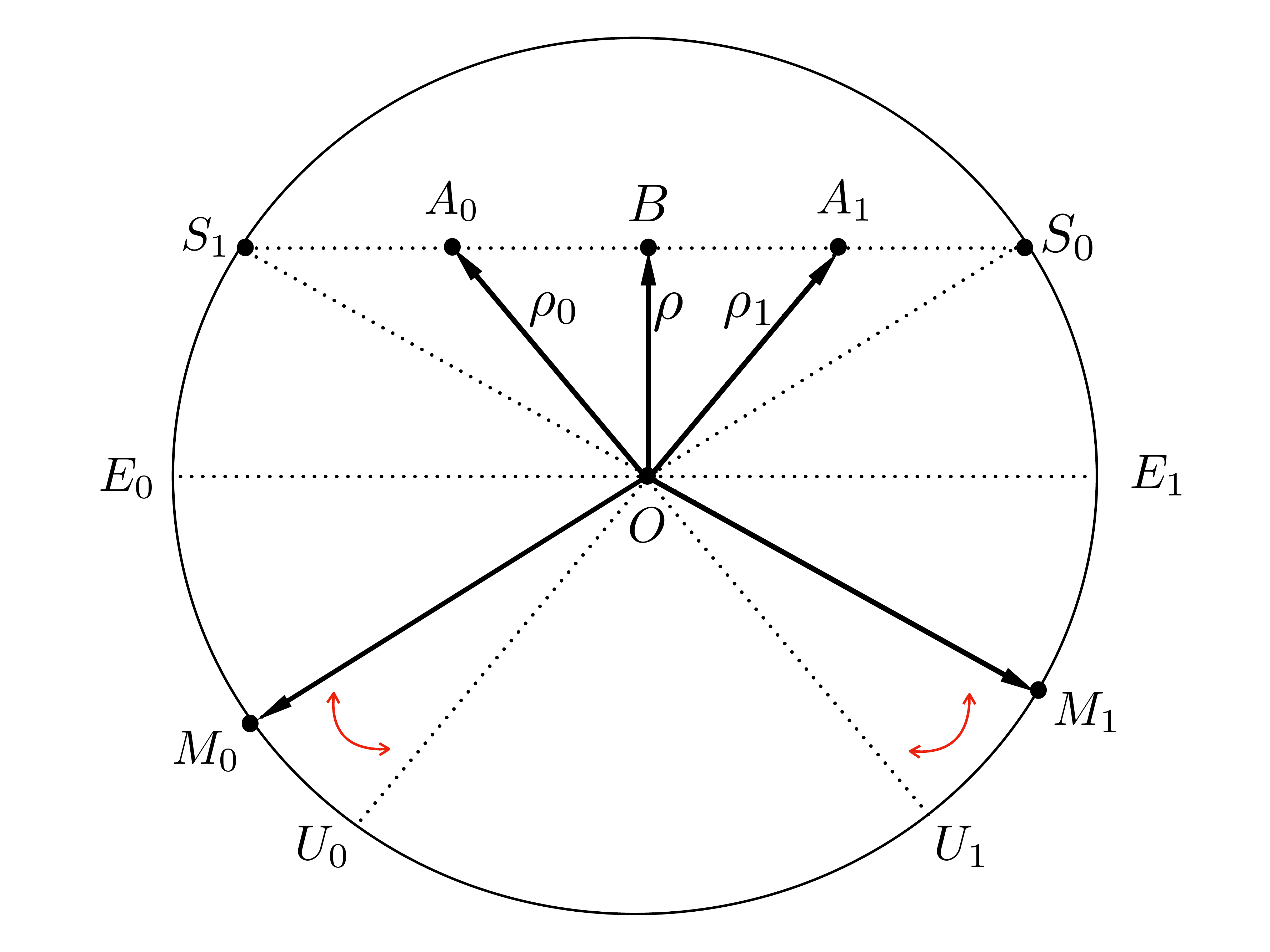}
  \caption{Two states in Eq. (\ref{eq:2s}) are depicted in the Bloch sphere. $OA$ and $OB$ denote the Bloch vectors of the states $\rho_0$ and $\rho_1$. Complementary states $\sigma_0$ and $\sigma_1$ on the sphere are found by extending $A_0 B$ and $A_1 B$ and consequently $OS_0$ and $OS_1$. Flipping them to their opposite directions, the MCM is obtained as $OM_0$ and $OM_1$, which coincides with $OU_0$ and $OU_1$ for $p=1$.
Note that when two states are pure, unambiguous discrimination is possible and is implemented by a POVM containing $OU_0$ and $OU_1$. The optimal measurement for minimum-error is given by $OE_0$ and $OE_1$. It is therefore found that the MCM, $OM_0$ and $OM_1$, is in-between unambiguous and minimum-error discrimination. }
  \label{fig:ex1}
\end{figure}

\subsection{ Two states}
The first example is two qubit states each prepared with equal {\it a priori} probabilities,
\bea
\rho_{\x} & = & p\ketbra{\psi_{\x}}+(1-p)\frac{\mathbbm{I}}{2},~~\x=0,1~~\mathrm{and} \label{eq:2s}\\
\rho & = &\frac{p}{2} (\ketbra{\psi_0}+ \ketbra{\psi_1})+(1-p)\frac{\mathbbm{I}}{2}. \nonumber 
\eea
Unambiguous discrimination is not possible for this ensemble states if $p<1$. Two pure states may be parameterized by $\cos\theta = \langle \psi_0 | \psi_1\rangle$, so we can without loss of generality write
\bea
\ket{\psi_{\x}} &=& \cos\frac{\theta}{ 2}\ket{0} + (-1)^{\x}\sin\frac{\theta}{2}\ket{1}. 
\eea
The maximum confidence for each state is computed as 
\bea
\max C(\x) = \frac{1}{2} (1+ \frac{p\sqrt{1- \cos^2\theta} }{ \sqrt{1-p^2 \cos^2\theta }}). 
\eea
The MCM can then be obtained from the Bloch vectors of the states:
\bea
\vec{n}(\rho_{\x})&&=((-1)^{\x} p\sin\theta,0,p\cos\theta)\\ 
\vec{n}(\rho)&&=(0,0,p\cos\theta).
\eea
From these, the Bloch vectors of the complementary states are found, using Eq. (\ref{eq:r}), to be
\bea
\hat{r}_{\x} & = & p( (-1)^{\x +1} t_{\x} \sin\theta, 0, \cos \theta ),\\ 
&& \mathrm{with}~~t_{\x} = \frac{\sqrt{1-p^2\cos^2\theta } }{ p\sqrt{1-\cos^2\theta }} 
\eea
where we note that $\|\hat{r}_{\x} \|=1$. An optimal POVM element is rank-one and can be described by a unit Bloch vector, denoted by $\hat{m}_{\x}$,\bea
\hat{m}_{\x} = - \hat{r}_{\x} & = & p( (-1)^{\x} t_{\x} \sin\theta, 0, -\cos\theta ). \label{eq:m2}
\eea
That is, an optimal POVM element for state $\rho_{\x}$ is given by $M_{\x} \propto (\mathbbm{I} + \hat{m}_{\x} \cdot \vec{\sigma})/2$ where $\sigma = (X,Y,Z)$ with Pauli matrices $X$, $Y$, and $Z$.

Remarks are in order. Firstly, suppose that pure states are given, i.e., $p=1$. Then, we have that $\hat{m}_{\x} = -\vec{n}(\rho_{\x+1})$, meaning $M_{\x}\perp \rho_{\x+1}$. In this case, an MCM coincides with unambiguous discrimination. Secondly, the MCM varies according to a noise parameter $p$, see Eq. (\ref{eq:m2}). Thirdly, for all values $p\in(0,1]$, an MCM is never a null measurement: the same holds true even if different {\it a priori} probabilities are given. That is, the act of not measuring can never give the maximal confidence. This contrasts with certain cases of minimum-error discrimination, in which a null measurement is optimal whenever $q_0- q_1 > \|q_0 \rho_0 -q_1 \rho_1 \|_1$, where $q_1$ and $q_2$ are {\it a priori} probabilities. 

 The probability of inconclusive outcomes can be minimized, see Eq. (\ref{eq:mphi2}). Since the {\it a priori} probabilities are equal, it is not difficult to see that $a_0=a_1$, from which it is straightforward to solve the optimization problem. It is obtained that the minimal probability of inconclusive outcomes is given by
\bea
Q = p | \langle \psi_0 | \psi_1\rangle |. 
\eea 
Note that the probability of inconclusive outcomes in USD is reproduced in Eq. (\ref{eq:usdinc}) with $p=1$.  

\subsection{ Geometrically uniform states}
A set of $N$ states $\{ \rho_{\x} \}_{\x=0}^{N-1}$ are geometrically uniform when there exists a unitary transformation $U$ such that $U\rho_{\x} U^{\dagger} = \rho_{\x+1}$ for all $\x$, i.e., $U^N=\mathbbm{I}$ \cite{1302298}. As one example, geometrically uniform qubit pure states can be written as
\bea
\ket{\psi_{\x}}=\cos\frac{\theta}{2}\ket{0}+e^{\frac{2 \pi i }{N} {\x}} \sin\frac{\theta}{2}\ket{1}
\eea
for some $\theta$. Note that a set of $N$ states $\{ \rho_{\x} \}_{\x=0}^{N-1}$ generalizes the three qubit states considered in Ref. \cite{croke2006}. Assume that the states are given with equal {\it a priori} probabilities: the ensemble is then given by 
\bea
\rho = \frac{1}{2} \left( \mathbbm{I} + (\cos\theta) Z \right)
\eea
 Since we consider pure states, we have $\mu_{\x} = 1/2$ for each $\x$, see Eq. (\ref{eq:f1}). The maximum confidence is given by 
\bea
\max C(\x) = \frac{2}{N}. 
\eea
In this case, it is shown that the maximum confidence concerning a particular state of interest only depends on the cardinality of an ensemble: a larger set shows a lower value of the maximum confidence and vice versa.

\begin{figure}[]
  \centering
  \includegraphics[scale=0.26]{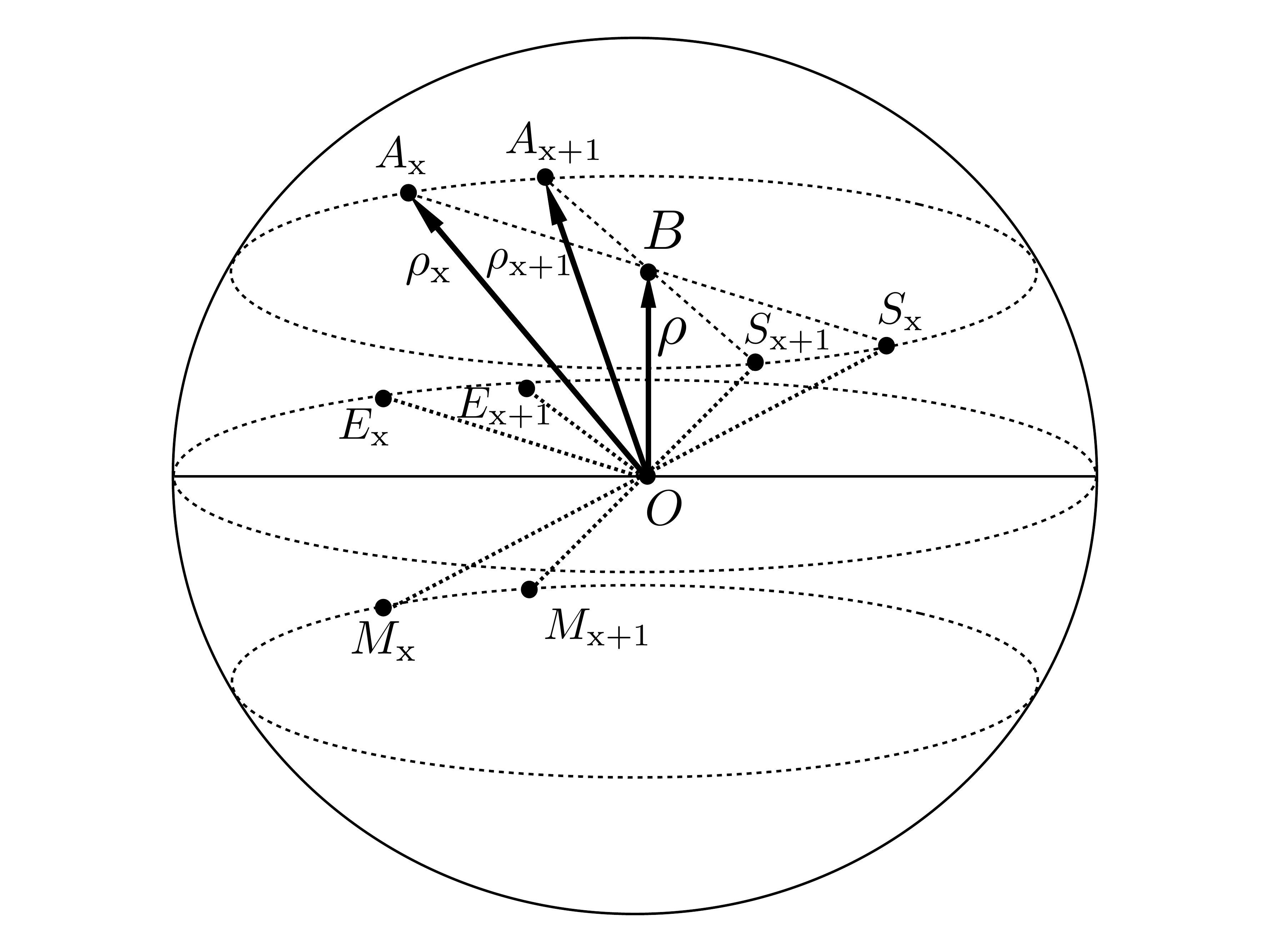}
  \caption{Geometrically uniform pure states $OA_{\x}$ form a circle defined by a radius $BA_{\x}$ where $OB$ denotes the ensemble of the states with equal probabilities. For a state $OA_{\x}$, its complementary state is found at $OS_{\x}$ by extending $BA_{\x}$. By rotating it with respect to the origin, an optimal POVM element $OM_{\x}$ that is orthogonal to the state $OS_{\x}$ is obtained.  The half-plane may be defined precisely by the collection of midpoints of $A_{\x}M_{\x}$. Note that the measurement for minimum-error discrimination contains $OE_{\x}$ obtained by projecting states $OA_{\x}$ onto the half-plane. Or, for those states in the half-plane an MCM coincides with a measurement for MED.}  
  \label{fig:ex2}
\end{figure}

An optimal measurement can be found as follows. Bloch vectors of the states are given by
\bea
\vec{n} (\rho_{\x}) & = & (\cos\frac{2\pi\x}{N} \sin\theta, \sin\frac{2\pi \x}{N}\sin\theta,\cos\theta) \\ 
\vec{n}(\rho) &= & (0,0,\cos\theta), 
\eea
and Bloch vectors of complementary states are obtained as
\bea
\vec{r}_{\x} & = &  ( -\cos\frac{2\pi\x}{N} \sin\theta, - \sin\frac{2\pi \x}{N}\sin\theta,\cos\theta) =- \hat{m}_{\x},~~ ~~~~~~
\eea
where $\{ \hat{m}_{\x}\}_{\x=1}^N$ denote Bloch vectors of optimal POVM elements. 

The minimal probability of inconclusive outcomes can be obtained by solving the optimization problem in Eq. (\ref{eq:mphi2}). Since the {\it a priori} probabilities are equal, it is not difficult to see that $a_0=a_1=\cdots=a_{N-1}$. Then, the minimal probability is computed as
\bea
Q = | \cos\theta|,
\eea 
which reproduces the result in Ref. \cite{croke2006} for the case of $N=3$.

Note that USD cannot be performed for for $N>2$ qubit states. The measurement for minimum-error discrimination is found on the half-plane, see Fig. \ref{fig:ex2}, and the guessing probability is given by $(1+\sin\theta)/N$ \cite{PhysRevA.87.012334}. That is, for geometrically uniform states with $\theta = \pi/2$, the MCM also performs minimum-error discrimination.

 \begin{figure}[t]
  \centering
  \includegraphics[scale=0.26]{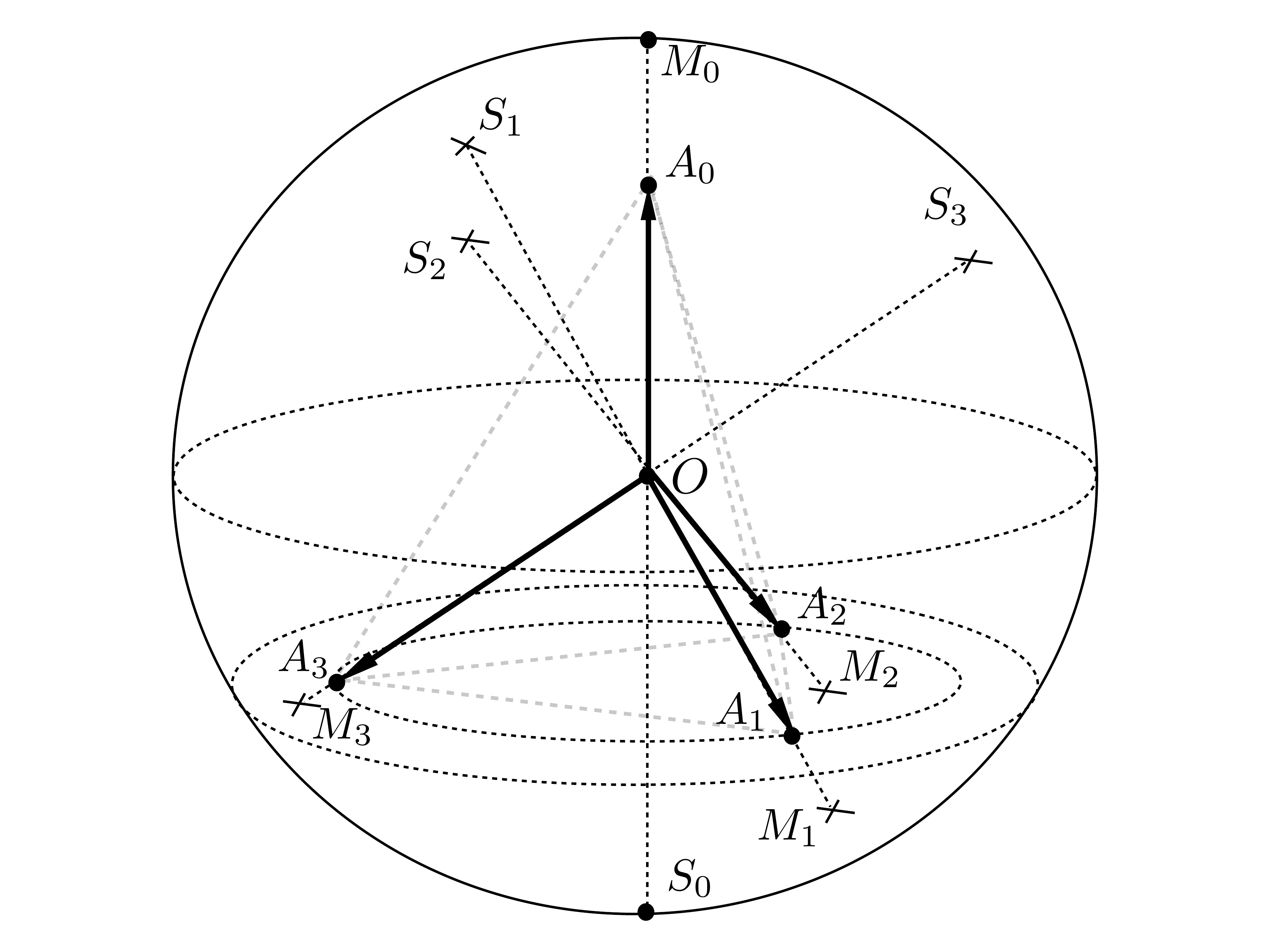}
  \caption{ The ensemble of noisy tetrahedron states $OA_{\x}$ for $x=0,1,2,3$ is given as $\mathbbm{I}/{2}$, corresponding to the origin $O$. Then, complementary states are found on the Bloch sphere in the opposite directions to given states: $OS_{\x}$ for $\x=0,1,2,3$. Optimal POVM elements are rank-one and found by rotating $OS_{\x}$ about $O$: $OM_{\x}$ for $\x=0,1,2,3$ are in the same direction to given states $OA_{\x}$.  }
  \label{fig:ex3}
\end{figure}

\subsection{ Tetrahedron states}
The next example we consider is an ensemble of tetrahedral states,
\bea
|\psi_{0}\rangle = |0\rangle,~\ket{\psi_{\x}}= \sqrt{ \frac{1}{3}} \ket{0} + e^{\frac{2\pi i \x}{3}} \sqrt{\frac{2}{3}}\ket{1},~\x=1,2,3.~~~~~~ 
\eea
so called because they form a tetrahedron in the Bloch sphere. These are symmetric, informationally complete (SIC) states, since $| \langle \psi_{\x} | \psi_{\y} \rangle | = 1/3$ for $\x\neq \y$ \cite{doi:10.1063/1.1737053}. 

To be more general, we consider noisy tetrahedron states 
\bea
\rho_{\x} = p | \psi_{\x} \rangle \langle \psi_{\x}| + (1-p) \frac{\mathbbm{I}}{2},~~\mathrm{for}~\x=0,1,2,3, 
\eea
given with equal {\it a priori} probabilities, so that $\rho = \mathbbm{I}/2$. From Eqs. (\ref{eq:f1}) and (\ref{eq:f2}), it follows that
\bea
\mu_{\x} = \frac{1}{1+p}~\mathrm{and}~ \max C(\x) = \frac{1+p}{4},~\mathrm{for}~\x=0,1,2,3.~~~~~~~
\eea
Note that for pure states the maximum confidence is given by $1/2$. Since the Bloch vectors of tetrahedron states are given by 
\bea
\vec{n}(\rho_{0}) & = & (0,0,p) \\
\vec{n}(\rho_{\x}) & = & (\frac{2\sqrt{2}}{3}p \cos\frac{2\pi \x}{3}, \frac{2\sqrt{2}}{3} p\sin\frac{2\pi \x}{3}, -\frac{1}{3}p ) \\
\vec{n}(\rho) & = & (0,0,0) , 
\eea
where $\x=1,2,3$, one can find the Bloch vectors of complementary states, 
\bea
\hat{r}_{\x} = { - \frac{1}{p} \vec{n}(\rho_{\x})} =  -\hat{m}_{\x},~~\mathrm{for}~\x=0,1,2,3. 
\eea
Thus, an MCM for tetrahedron states is shown in Fig. \ref{fig:ex3}. It is worth mentioning that the MCM coincides with the minimum-error measurement for the tetrahedron states \cite{bae2013, PhysRevA.87.012334}. We also remark that, as it is shown in \ref{subsec:inc}, since the convex hull of the projectors contains the identity, an MCM can be constructed such that inconclusive outcomes do not occur.

\begin{figure}[]
  \centering
  \includegraphics[scale=0.26]{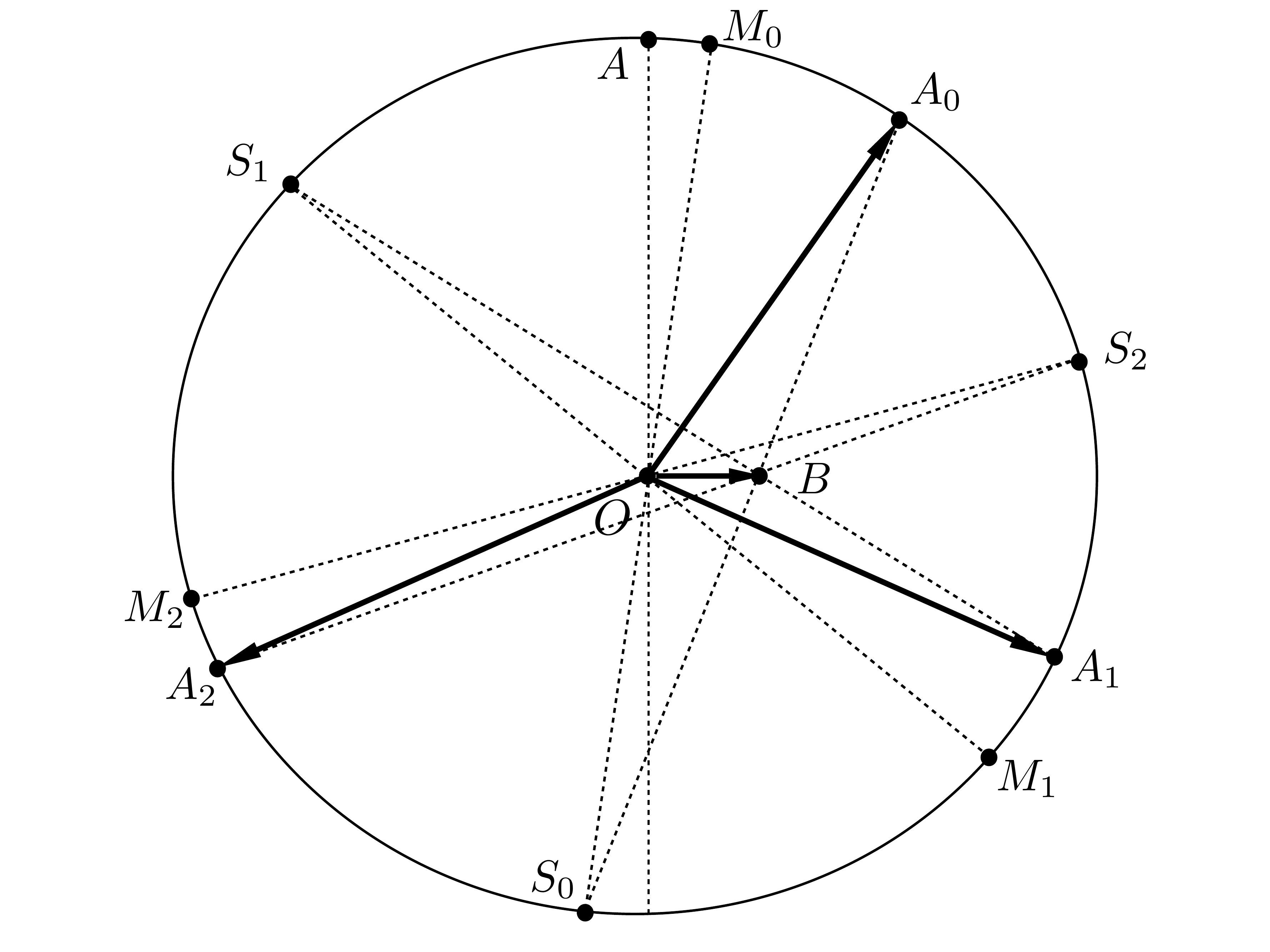}
   \caption{Three states $OA$, $OA_1$ and $OA_2$ are geometrically uniform. The first state is slightly tilted so that an ensemble of three states $OA_{\x}$ for $\x=0,1,2$ is considered. The ensemble is denoted by $OB$. Complementary states $OS_{\x}$ are found by extending $A_{\x} B$, and optimal POVM elements are found by inverting $OS_{\x}$ with respect to $O$. None of the elements $OM_{\x}$ are identical to states $OA_{\x}$. }
  \label{fig:ex4}
\end{figure}

\subsection{ Asymmetric states I }

In this subsection, we consider the ensemble of three asymmetric states, which is constructed by slightly modifying one of the three geometrically uniform states. We look at the three states
\bea
&& \ket{\psi_0} = \cos\frac{\theta}{2}\ket{0}+\sin\frac{\theta}{2}\ket{1},~~\mathrm{and}~~ \nonumber \\
&& \ket{\psi_1} = \frac{1}{2}\ket{0}+\frac{\sqrt{3}}{2}\ket{1}),~ \ket{\psi_2} = \frac{1}{2}(\ket{0}-\sqrt{3}\ket{1}). \label{eq:asy1}
\eea 
That is, two states $|\psi_1\rangle$ and $|\psi_2\rangle$ are fixed and a state $|\psi_0\rangle$ is varied by an angle $\theta$. The Bloch vectors are 
\bea
\vec{n}(\psi_0)&=&(\sin\theta,0,\cos\theta),~~ \vec{n}(\psi_1) = (\frac{\sqrt{3}}{2},0,-\frac{1}{2}) \\
\vec{n}(\psi_2)&=&(\frac{-\sqrt{3}}{2},0,-\frac{1}{2}), ~~\vec{n}(\rho)=\frac{1}{3}(\sin\theta,0,-1+\cos \theta). \nonumber 
\eea
It turns out that an MCM for them does not contain any symmetry, as can be seen in Fig. \ref{fig:ex4}. We make use of the expression in Eq. (\ref{eq:ct}) to get
\bea
t_0 &=&\frac{9 - 2 (1-\cos\theta )}{ 9 - 4(1-\cos \theta )} ~~\mathrm{and} \\
t_{\x}&=&\frac{ 9 - 2 (1-\cos\theta)}{ 9- (1-\cos \theta ) + 3 \sqrt{3} (-1)^{\x} \sin\theta},~\x=1,2.~~~ 
\eea
The maximum confidence is found to be $(1+1/t_{\x})/3$. It is seen that an MCM for the asymmetric states does not contain any symmetry. Since the convex hull of the projectors contains the identity, an MCM without inconclusive outcomes can be constructed.

\begin{figure}[]
  \centering
  \includegraphics[scale=0.26]{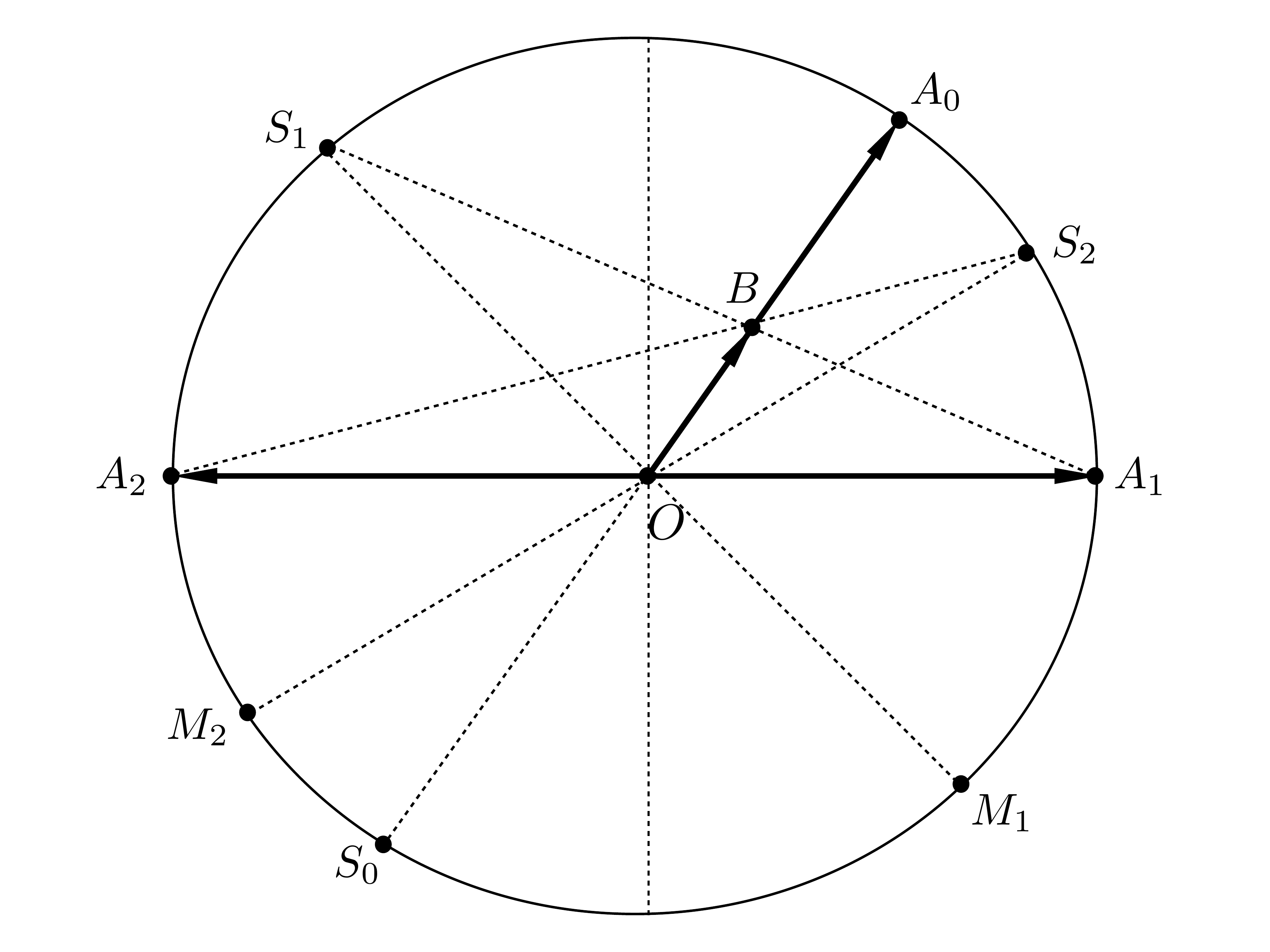}
  \caption{Three states $OA_0$, $OA_1$ and $OA_2$ are considered where $OA_1$ and $OA_2$ are orthogonal. Complementary states $OS_{\x}$ are found on the sphere by extending $A_{\x}B$. An optimal POVM consists of $OA_0$, $OM_1$, and $OM_2$. A measurement for minimum-error discrimination contains two POVM elements $OA_1$ and $OA_2$. }
  \label{fig:ex5}
\end{figure}

\subsection{ Asymmetric states II}

The next example of asymmetric states considered is the following
\bea
\ket{\psi_0} = \cos\frac{\theta}{2}\ket{0}+\sin\frac{\theta}{2}\ket{1},~~\ket{\psi_1} = \ket{+}, ~\mathrm{and}~\ket{\psi_2} = \ket{-} \nonumber
\eea
where each state is prepared with equal {\it a priori} probability. Similarly to the previous case, two states $|\psi_1\rangle$ and $|\psi_2\rangle$ are fixed and a state $|\psi_0\rangle$ varies by an angle $\theta$. In contrast with the ensemble in Eq. (\ref{eq:asy1}), the pair of states $|\pm\rangle$ contains a symmetry: they are invariant under a rotation about the $x$-axis. Their Bloch vectors are 
\bea
\vec{n}(\psi_0) &=&(\sin\theta,0,\cos\theta),~~\vec{n}(\psi_1) = (1,0,0) \nonumber \\
\vec{n}(\psi_2) &=&(-1,0,0),~~ \vec{n}(\rho) = \frac{1}{3}(\sin\theta,0,\cos\theta). ~~
\eea
We again exploit the expression in Eq. (\ref{eq:ct}) to find 
\bea
t_0 = 2, ~~ t_1 = \frac{4}{5- 3\sin\theta},~~t_2 = \frac{4}{5+3\sin\theta}. ~~
\eea
It follows that 
\bea
\max C(\x) = \frac{1}{3}(1+\frac{1}{t_{\x}}). 
\eea
Interestingly, the maximum confidence for the state $|\psi_0\rangle$, which is parameterized by $\theta$, does not depend on the angle. The maximum confidence for the other two states depends upon the angle $\theta$ from the other state $|\psi_0\rangle$.

In contrast to the three states in the case of the ensemble in Eq. (\ref{eq:asy1}), the MCM contains a symmetry, seen from the Bloch vectors of the complementary states which are
\bea
\hat{r}_0 &=&- \vec{n} (\psi_0)\\ 
\hat{r}_1 &=&(\frac{1}{3}\sin\theta-1,0,\frac{1}{3}\cos\theta)t_1+\frac{1}{3}(\sin\theta,0,\cos\theta) \nonumber \\
\hat{r}_2 &=&(\frac{1}{3}\sin\theta+1,0,\frac{1}{3}\cos\theta)t_2+\frac{1}{3}(\sin\theta,0,\cos\theta). \nonumber
\eea
That is, an optimal POVM element for the state $|\psi_0\rangle$ shares its Bloch vector with the state $\vec{n}(\psi_0)$. An MCM for two states $|\pm\rangle$ depends on the angle $\theta$ of the other state $|\psi_0\rangle$. Since the convex hull of the projectors of an MCM for the asymmetric states contains the identity, the probability of inconclusive outcomes is also zero. 

In the case of minimum-error discrimination for the ensemble, an optimal measurement does not aim to detect a state $|\psi_0\rangle$. It contains two POVM elements having Bloch vectors $\vec{n}(\psi_1)$ and $\vec{n}(\psi_2)$. Then, a detection event on the first (second) POVM element characterized by $\vec{n}(\psi_1)$ ($\vec{n}(\psi_2)$) concludes a state $|\psi_1\rangle$ ($|\psi_2\rangle$). In this way, the guessing probability is given as $2/3$ \cite{PhysRevA.87.012334}.



\section{Conclusion}
\label{sec:con}

In summary, we have investigated MCMs for qubit states. We have presented a simple scheme to find MCMs for qubit states when an ensemble and a state of interest are given. The scheme exploits the geometry in a Bloch sphere without resorting to the computational optimization problem. We then considered various qubit states. From the cases of two qubit states, it is shown that an MCM lies between two strategies, minimum-error and unambiguous discrimination. An MCM for geometrically uniform states generalizes an example from Ref. \cite{croke2006}. An MCM for tetrahedron states is identical to a measurement for minimum-error discrimination. Otherwise, when an ensemble does not contain any symmetry, it was seen that MCMs highly depends on the particular state of interest. 

Our results elucidate the meanings of different qubit measurements, each of which may aim to maximize different figures of merit. Measurements for various qubit ensembles may also be used to devise quantum protocols to certify the properties of qubit states.


\section*{Acknowledgement}
KF, HL, and JB were supported by National Research Foundation of Korea (NRF-2021R1A2C2006309, NRF-2022M1A3C2069728), Institute of Information \& communications Technology Planning \& Evaluation (IITP) grant (the ITRC Program/IITP-2021-2018-0-01402). JBB and CRC were supported by the Independent Research Fund Denmark and a KAIST-DTU Alliance stipend.


%

\end{document}